\documentclass [12pt]{amsart}
\usepackage{amscd}
\usepackage{amsxtra}
\usepackage{amsthm}
\usepackage{epsfig}
\usepackage{graphicx,color}
\usepackage{amssymb}
\usepackage[all]{xy}
\usepackage{mathbbol}
\usepackage[ansinew]{inputenc}
\usepackage[T1]{fontenc}
\newtheorem{theorem}{\sc Theorem}
\newtheorem{proposition}{\sc Proposition}
\newtheorem*{lemma}{\sc Lemma}

\theoremstyle{remark}
\newtheorem{remark}{\it Remark}
\newtheorem{example}{\it Example}
\newtheorem{definition}{\sc Definition}

\font\tmsb=msbm10 at12pt
\font\smsb=msbm7
\font\ssmsb=msbm5
\newfam\msbfam
\textfont\msbfam=\tmsb
\scriptfont\msbfam=\smsb
\scriptscriptfont\msbfam=\ssmsb
\def \1{\mathbb {1}}
\def \RM{\mathbb {R}}
\def \CM{\mathbb{C}}

\def \KM {\mathbb{K}}

\def \HM {\mathbb{H}}

\def \Ct {\mathcal{C}}

\def \Spec {{\rm Spec\,}}
\def \Der {{\rm Der\,}}
\def \Aut {{\rm Aut\,}}
\def \exp {{\rm exp\,}}
\def \Id {{\rm Id\,}}
\def \d{\partial}

\def\dt{\delta} 
\def\a{\alpha}
\def\b{\beta}

\def\l{\lambda}

\def\p{\varphi}  
\def\lb{\left\{}
\def\rb{\right\}}
\def\G{\Gamma}

\def \s{\sigma}
\def \t{\tilde}

\def \to{\longrightarrow} 
\def \w{\wedge}

\def \< {{\langle }}
\def \> {{\rangle }}

\newcommand{\C}{{\mathbb C}}

\newcommand{\iso}{\approx}

\newcommand{\It}{{\mathcal I}}

\newcommand{\Ot}{{{\mathcal O} }}

\begin{document}
\title [Classical and quantum integrability]{Classical and  quantum integrability}

\author{Mauricio D. Garay and Duco van Straten}
\date{February 2008}
\address{Duco van Straten}
\address{ Fachbereich 8, Institut f\"ur Mathematik,
Staudingerweg 9, Johannes Gutenberg-Universit{\"a}t,
55099 Mainz, Germany.}
\email{straten@mathematik.uni-mainz.de}
\address{Mauricio D. Garay}
\address{IH\'ES, Le Bois Marie, 35 route de Chartres, Bures sur Yvette, France.}
\address{Guest of the SFB TR 45, Fachbereich 8, Institut f\"ur Mathematik, Staudingerweg 9, Johannes Gutenberg-Universit{\"a}t,
55099 Mainz, Germany.}
\email{garay@ihes.fr}



\parindent=0cm


\begin{abstract}{It is a well-known problem to decide if a classical hamiltonian 
system that is integrable in the Liouville sense, can be quantised to a quantum 
integrable system. We identify the local obstruction to do so and show that it
vanishes under certain conditions.}
\end{abstract}
\maketitle
\section*{Introduction}
Let $M$ be an open subset of the complex symplectic manifold  $T^*\CM^n \approx \CM^{2n}$ and consider
an {\em integrable system} $f: M \to S$, i.e., the fibres of the morphism $f$ are of pure dimension $n$ and the restriction of the symplectic form to the smooth locus of any fibre vanishes. If we take local coordinates on the base, an integrable system is given by holomorphic functions
$f_1,\dots,f_k:M \to \CM$
whose Poisson brackets vanish pairwise: 
$$\{ f_i,f_j \}=0,\;\;1 \le  i,j \le k$$
and any other holomorphic function commuting with the $f_i$  is a function of them.
A basic question is the following: does there exist a quantum integrable system whose classical limit is
the given classical integrable system? In other words, we ask for the existence of  {\em commuting} $\hbar$-differential 
operators $F_1,\dots,F_k$ whose principal symbols are the given $f_1,\dots,f_k$.
This question makes sense both in the algebraic, in the holomorphic and in the real $C^\infty$ setting.\\
In this paper, we attach to $f$ a certain  complex  $C_f^\cdot$ on $M$, together with
a certain {\em anomaly classes} $\chi \in H^2(C_f^\cdot )$ that are obstructions to extend the
quantisation to the next order in $\hbar$. The quantisation problem can be solved provided that 
all these classes vanish \cite{vanStraten_quantique}.\\
 Because of the close relation between the complex $C_f^\cdot$ and the relative de Rham complex $\Omega_f^\cdot$, the anomaly class is of {\em topological nature} and vanishes under reasonable topological conditions on the map $f$.\\
The anomaly classes can be also defined for general involutive systems but in that case, we were not able to prove any quantisation property.
There is a considerable literature on the subject of quantisation, going back to early days of quantum mechanics, but we will not try to give a complete overview in this paper.
Quantum integrability is regarded sometimes as a remarkable fact and sometimes -based on heuristic arguments- as the general rule \cite{Hietarinta1,Robnik}. It seems that the results and the constructions of this paper can be extended to Poisson manifolds but we did not check all details.
\section{The quantisation theorem}
\subsection{The Heisenberg quantisation.}
The deformation $Q=R[[\hbar]]$ of the polynomial ring $R=\CM[q_1,\dots,q_n,p_1,\dots,p_n]=\CM[q,p]$
is an associative algebra for the   {\em normal product}
$$f \star g:=e^{\hbar \sum_i \d_{p_i} \d_{q_i'}}f(q,p) g(q',p')_{|(q=q',p=p')}$$
and defines a flat deformation of $R$ over $\CM[[\hbar]]$ \cite{Moyal}.
The relation
$$p \star q-q \star p=\hbar $$
shows that the map which sends $p$ to $\hbar \d_x$ and $q$ to the multiplication by $x$ induces an isomorphism of this algebra with the algebra $\CM[x,\hbar \d_x][[\hbar]]$
of differential operators generated by $\hbar \d_x$ and $x$ depending on a parameter $\hbar$.\\
More generally, one can consider a field $\KM$ and an algebra $A$ which is a {\em flat deformation} over $\KM[[\hbar]]$
of a commutative ring $B$. This means that $A$ has the structure of a $\KM[[\hbar ]]$-algebra,
such that
\begin{enumerate}
\item $A/\hbar A =B$,
\item $A=\underleftarrow{\lim}A/\hbar^l A $
\item multiplication by the central element $\hbar$ is injective.
\end{enumerate}
Such an algebra will be called a {\em quantisation} of the ring $B$. The quantisation $Q$ described above will be called the
{\em Heisenberg quantisation} of $R$.\\
This situation is summarised by the diagram
$$\xymatrix{A\ar[r]^\hbar & A \ar[r] & B\\ 
            \KM[[\hbar]] \ar[r]^\hbar \ar[u]&  \KM[[\hbar]] \ar[r] \ar[u]& \KM \ar[u] } $$
The formula for the normal product defines Heisenberg quantisation for the following rings in a similar way:
\begin{enumerate}
\item $R_{an}=\CM\{q_1,\dots,q_n,p_1,\dots,p_n \}$ the ring of holomorphic function germs at the origin in $T^*\CM^n$ ,
\item $\G(U,\Ot_{T^*\CM})$ the ring of holomorphic function in an open subset $U \subset T^*\CM^n$,
\item $R_{\infty}=C^\infty_{2n}$ the ring of $C^\infty$ function germs at the origin in $T^*\RM^n$,
\item $\G(U,C^\infty_{T^*\RM^n})$ the ring of $C^\infty$ function germs in an open subset $U \subset T^*\RM^n$.
\end{enumerate}
These rings are stalks or global section of sheaves, the notion of quantisation admits a straightforward variant
for sheaves.\\
These notions are of course classical, going back to the early days of quantum mechanics when Born, Heisenberg, Jordan and Dirac
proposed to replace the commutative algebra of hamiltonian mechanics by the non-commutative one over the Heisenberg algebra \cite{BHJ,Dirac} (see also \cite{source_book}).
The approach of star products was introduced in the more general context of symplectic manifolds
by Bayen-Flato-Fronsdal-Lichnerowicz-Sternheimer \cite{Flato_al} (see also \cite{Fedosov}).
The link between both approaches was re-phrased into modern terminology by Deligne \cite{Deligne_star}.
\subsection{The quantisation problem}
\label{SS::def}
We consider the following problem:\\
Let $A$ be a quantisation of a ring $B$ and let $f_1,\dots,f_k$ be elements in the ring $B$.
Under which condition can we find {\em commuting} elements $F_1,\dots,F_k$ in $A$ such that $F_i=f_i \mod \hbar$?\\

In such a situation, we call $F_1,\dots,F_k $ a {\em quantisation} of $f_1,\dots,f_k$.
From the point of view of naive quantum mechanics, this would mean if it is possible to measure simultaneously the quantities $F_1,\dots,F_k$.\\
There is an obvious obstruction to perform a quantisation of $f_1,\dots,f_k$ that we shall now explain.\\
The canonical projection  $\sigma: A \to A/\hbar A =B$ is called the {\em principal symbol}.
The result of commuting two elements $F,G \in A$ is divisible by $\hbar$ and its class $\mod \hbar^2$
only depends on the symbols $f=\sigma(F)$ and $g=\sigma(G)$. In this way, one obtains a well-defined 
Poisson algebra structure $\{-,-\}$ on $B$ by putting
\[  \{f,g\}:=\frac{1}{\hbar}[F,G] \mod \hbar \]
Recall that this means that this bracket satisfies the Jacobi-identity and is a derivation on $B$ in both variables.
For the Heisenberg quantisation, we get the standard
formula
$$\{ f , g \}=\sum_{i=1}^n \d_{p_i}f \d_{q_i}g- \d_{q_i}f \d_{p_i}g .$$
For the answer to our question to be positive, we need that the Poisson brackets of the $f_i$'s vanish. This justifies the following definition.
\begin{definition}A collection of elements $f=(f_1,\dots,f_k)$ of a Poisson algebra $B$ is called an {\em involutive system} if the elements
Poisson-commute pairwise.
\end{definition}
Our question can be made more precise:\\
Let $A$ be a quantisation of a ring $B$ and let $f=(f_1,\dots,f_k)$ be an involutive system  in the ring $B$.
Under which conditions can we find commuting elements $F_1,\dots,F_k$ in $A$ such that $F_i=f_i \mod \hbar$?\\

We shall provide an answer only for integrable system. For the symplectic case, we mean the following.
The choice of elements $f_1,\dots,f_k \in B$ (involutive or not) induces in $B$ a $\KM[[t]]:=\KM[[t_1,\dots,t_k]]$-algebra structure on $B$ defined by
$$t_im:=f_im. $$

\begin{definition}
An involutive system $f=(f_1,\dots,f_n),\ f_i \in R=\CM[q_1,\dots,q_n,p_1,\dots,p_n]$ is called an {\em integrable system}
if it defines a flat $\KM[[t]]$-algebra.
\end{definition}
We will formulate a sufficient condition for the quantisation of integrable systems.
\subsection{The complex $C_f^\cdot$}
Let $B$ be a commutative Poisson algebra over a field $\KM$ and let $f=(f_1,\dots,f_k)$
be an involutive system. Consider the complex with terms
\[ C_f^0=B,\ C_f^p:=\bigwedge^p B^k \]
and the $\KM$-linear differential  $\dt:C_f^p \to C_f^{p+1}$ is defined by
$$ \dt(m\, v)=\sum_{j=1}^k\{f_j,m \} v \wedge e_j, $$
where  $e_1=(1,0,\dots,0)$, \dots, $e_k=(0,\dots,0,1)$ denotes the canonical basis in $B^k$,
$v \in \bigwedge^p B^k$ and $m \in B$.\\
Note that the differential in the complex $(C_f^\cdot,\dt)$ is $\KM[t]$-linear and
not $B$-linear. As a result, the above cohomology groups have the structure of  $\KM[t]$-modules defined by
$$t_i[m]:=[f_im] $$
where $[m]$ denotes the cohomology class of the coboundary $m$. The cohomology module of the complex $(C_f^\cdot,\dt)$
will be denoted by $H^p(f)$.
\subsection{Statement of the theorem}
 The main result of this paper is the following theorem.
\begin{theorem}
\label{T::quantisation}
Let $f=(f_1,\dots,f_n), f_i \in R=\CM[q_1,\dots,q_n,p_1,\dots,p_n]$ be an integrable system.
If the module $H^2(f)$ is torsion free then the integrable system $f$ is quantisable, i.e.,
there exists commuting elements $F_1,\dots,F_n \in Q=R[[\hbar]]$ such that $F_i=f_i \ ({\rm mod}\ \hbar)$.
\end{theorem}
Analogous results hold for the rings $R_{an},R_{\infty},\G(U,\Ot_{T^*\CM^n}),\G(U,C^\infty_{T^*\RM^n})$,
if we assume that the map $f$ satisfies the $a_f$-condition \cite{Thom_strates}.
The module $H^2(f)$ is torsion free in all cases known to the authors, this will be part of a forthcoming paper.

\section{Anomaly classes and topological obstructions}
\subsection{Cohomological obstruction to quantisation}
Let us come back to the general problem of quantising an involutive system given a quantisation $A$ of a ring $B$.\\
We have seen that a quantisation of $B$ induces a Poisson algebra structure on it.
The $\KM[[\hbar]]$-algebra $A$ is itself a {\em non-commutative} Poisson algebra, the Poisson bracket being defined by the formula
\[ \{F,G\}=\frac{1}{\hbar}[F,G] .\]
In fact, the non-commutative algebras obtained by higher order truncations
\[A_l:=A/\hbar^{l+1} A,\ l \geq 0\]
also admit Poisson algebra structures.
The Poisson bracket in $A_l$ is defined by
\[ \{\sigma_l(F),\sigma_l(G)\}=\s_l(\frac{1}{\hbar}[F,G]) \]
where $\s_l:A \to A_l$ denotes the canonical projection.
In the sequel, we abusively denote these different Poisson brackets in the same way.\\
From the flatness property, one obtains exact sequences
\[ \xymatrix{ 0 \ar[r] & B  \ar[r] & A_{l+1} \ar[r] & A_l \ar[r] & 0 }\]
induced by the identification
\[\hbar^{l+1}A_{l+1} \iso B \hbar^{l+1}/\hbar^{l+2} \iso B. \]
We will use this identification without further mention.\\
We try construct quantisations of an involutive system $f=(f_1,\dots,f_k)$ order for order in $\hbar$.
\begin{definition}{An $l$-lifting of an involutive system $f=(f_1,\dots,f_k)$, $f_i \in B$
is a collection of Poisson commuting elements $ F=( F_1,\dots, F_k)$, $F_i \in  A_l$,
such that the principal symbol of $F_i$ is $f_i$.}
\end{definition}

Consider an arbitrary $l$-lifting  $F$ of our involutive mapping $f$.
Take any elements $G_1,\dots,G_k \in A_{l+1}$ which project to $F_1,\dots,F_k$.
As the $F_i$'s Poisson commute in $A_l$, we have 
\[ \{ G_i,  G_j \}=\chi_{ij}\hbar^{l+1} \]
\begin{proposition}
\label{P::cocycle}
\ \\
\begin{enumerate}
\item The element $\chi(G):=\sum \chi_{ij}e_i \w e_j$ defines a 2-cocycle in the complex $C_f^\cdot$, 
\item its cohomology class $\chi(G)$ depends only on the $l$-lifting
$F$ and not on the choice of $G $. 
\end{enumerate}
\end{proposition}
\begin{proof}
Write
$$ \chi(G)=\sum_{i,j \geq 0}\chi_{ij}e_i \w e_j,\ \chi_{ij} \in B $$
with $\hbar^{l+1} \chi_{ij}=\{ G_i,  G_j \} $.\\
We have
$$\dt \chi(G)=\sum_{i,j,l \geq 0}v_{ijl}e_i \w e_j \w e_l $$
with $\hbar^{l+1}v_{ijl}=\hbar^{l+1}\{\chi_{ij},f_l\}=\{ \{ G_i,  G_j \},G_l \} $.
Therefore the Jacobi identity implies that $v_{ijl}+v_{lij}+v_{jli}=0$. This proves that $\chi$ is a cocycle.\\
Now, take $\t G_1,\dots,\t G_k \in A_{l+1}$ which also project to $F_1,\dots,F_l$, then  $\t G_j=G_j+\hbar^{l+1} m_j$ for some
$m_1,\dots,m_k \in B$. Consider  the 2-cocycle $ \chi(\t G)=\sum \t \chi_{ij} e_i \w e_j$ associated to $\t G$.
One then has:
$$\{ G_i+\hbar^{l+1} m_i,G_j+\hbar^{l+1} m_j \}=\chi_{ij}\hbar^{l+1}
+\hbar^{l+1}(\{ f_i,m_j \}+\{m_i,f_j \}).$$
We get the equality
$\chi(\t G)=\chi(G)+\dt(m)$ where $m=\sum_{i=1}^k m_i e_i=(m_1,\dots,m_k)$, therefore the cohomology class
of $\chi(G) $ depends only on $F$ and not on $G$.  This concludes the proof of the proposition.
\end{proof}

\begin{definition}{The cohomology class $\chi_F:=[\chi(G)] \in H^2(f)$ 
is called the {\em anomaly class} associated to
the $l$-lifting $F$.}
\end{definition}
Summing up our construction one has the following result.
\begin{proposition}
Let $f=(f_1,\dots,f_k)$ be an involutive map.
An $l$-lifting $F=(F_1,\dots,F_k)$, $F_i \in A_l$ of $f$
extends to an $(l+1)$-lifting if and only if the anomaly class 
$\chi_F \in H^2(f)$ vanishes.
\end{proposition} 
\begin{proof}
Take $ G_1,\dots, G_k \in A_{l+1}$ which project to $F_1,\dots,F_k \in A_l$.
The anomaly class $\chi_F \in H^2(f)$ vanishes, thus there exists
$m \in R^n$ such that $\chi(G)=\dt m$ with $m=\sum m_i e_i=(m_1,\dots,m_k)$.
The map $G-\hbar^{l+1}m$ is an $(l+1)$-lifting of $f$.\\
Conversely assume that the $l$-lifting $F$ admits an $(l+1)$-lifting $G$, then $\chi(G)=0$ and a fortiori
its cohomology class vanishes.
\end{proof}
\subsection{Relation with the de Rham complex}
Consider the case where $f=(f_1,\dots,f_n)$ is an integrable system in the ring $R=\CM[q,p]$.\\
The ring $R$ has a $T=\CM[t]$-module structure defined by $t_ia:=f_ia $, $a\in R$.
The de Rham complex $\Omega^\cdot_{R/T} $ is defined by $\Omega_{R/T}^0=R$, $\Omega^1_{R/T}=\Omega^1_R/f^*\Omega^1_T$,
$\Omega^k_{R/T}=\bigwedge^k \Omega^1_{R/T} $ and the differential is the usual exterior differential.\\
An element of $\Omega^k_{R/T}$ is an algebraic $k$-form in the $q,p$ variables defined modulo forms of the type
$df_1 \w \a_1+\dots+df_n \w \a_n$:
$$\Omega_{R/T}^k:=\Omega_R^k/f^*\Omega^1_T \w \Omega^{k-1}_R, {\rm for\ } k>0,\ \Omega_{R/T}^0=R. $$
We show the existence of mapping of complexes
$$(\Omega^\cdot_{R/T},d) \to (C^\cdot_f,\dt) .$$
The interior product 
$$v \mapsto i_v \omega $$
of a vector field $v$ with the symplectic form $\omega= \sum_{i=1}^n dq_i \w dp_i$
induces an isomorphism between the space of one-forms $\Omega^1_{R}$ and that of vector fields $\Der(R,R)$.
If the form is closed then the corresponding vector is called {\em locally hamiltonian}, if the one-form is exact
it is called a {\em hamiltonian vector field}.\\
 The {\em hamiltonian vector field associated to a function} $H$ is the field associated to $dH$, it is given by the formula
$$\sum_{i=1}^n \d_{p_i}H \d_{q_i}-\d_{q_i}H \d_{p_i}. $$
Denote by $v_1,\dots,v_n$ the hamiltonian vector fields of the functions $f_1,\dots,f_n$, the mapping
$$\Omega^1_{R/T}  \to C^1_f,\ \a \mapsto (i_{v_1}\a,\dots,i_{v_n}\a )  $$
induces a mapping
$$\p^k:\bigwedge^k\Omega^1_{R/T}=\Omega^k_{R/T} \to \bigwedge^kC^1_f=C^k_f.$$ It is readily checked
that these maps commute with differentials, therefore we get  a map complexes $\p^\cdot:(\Omega^\cdot_{R/T},d) \to (C^\cdot_f,\dt)  $.
\subsection{Topological anomalies}
An integrable system $f=(f_1,\dots,f_n)$ induces a morphism
$$f:M=\Spec(R)=\CM^{2n} \to S=\Spec(T)=\CM^n .$$
The relative de Rham complex and the complex  $C^\cdot_f$ can both be sheafified to complexes
$(\Ct^\cdot_f,\dt)$, $(\Omega_f^\cdot,d)$  on the affine space $M$.
As $M$ is affine, there is an isomorphism
$$H^2(f) \approx \HM^2(\CM^{2n},\Ct^\cdot_f) $$
induced by the vanishing of higher cohomology groups for the coherent sheaves.\\
The map $\p^\cdot$ of the previous subsection extends to a map of sheaf-complexes that we denote by the same symbol.
By a direct local computation, we get the following result.
\begin{proposition}
\label{P::deRham}
The map $\p^\cdot :(\Omega_f^\cdot,d) \to (C_f^\cdot,\dt)$ is an isomorphism of sheaves
at the smooth points of the morphism $f$.
\end{proposition}
The direct image sheaf $f_*\CM_M$ is constructible and defines a (maximal) Zariski open subset $S'$ over which $f$ is smooth and \cite{Grothendieck_deRham}
$$\HM^k(S',f_*\Omega^\cdot_f) \approx (\RM^k f_*\CM_{X})_{\mid S'} \otimes \Ot_{S'}.$$
Proposition \ref{P::deRham} gives an isomorphism
$$\HM^k(S',f_*\Ct^\cdot_f) \approx \HM^k(S',f_*\Omega^\cdot_f). $$
There is a chain of maps (where we use the notation $\sim$ for isomorphisms)
$$\xymatrix{H^k(f) \ar[r]^-{\sim} &\HM^k(M,\C^\cdot_f) \ar[r]^-{\sim}&
\G(S,\RM^k f_*\Ct^\cdot_f) \ar`r[d]`d[l]`[lld]_-r[lld]\\
\G(S',\RM^k f_*\Ct^\cdot_f) \ar[r]^-{\sim}& \G(S',\RM^k f_*\Omega^\cdot_f) \ar[r]^-{\sim}& \G(S',\RM^k f_*\CM_{X})} $$
where $r$ denotes the restriction.\\
Conclusion: To each element of $H^k(f)$ is associate a section of the topological cohomology bundle
$\bigcup H^k(f^{-1}(s),\CM) \to S'$.\\
The image of an anomaly class $\chi_F$ will be a called the associated {\em topological anomaly}.
\begin{example}Take $n=1$, $f=pq$, the complexes $(\Ct^\cdot_f,\dt)$, $(\Omega_f^\cdot,d)$ have both two terms
$$\xymatrix{\Ot_M \ar[r]^d \ar[d]^-{\p_0=\Id} & \Omega^1_f \ar[d]^-{\p^1} \\
             \Ot_M \ar[r]^\dt  & \Ct^1_f \approx \Ot_M.}$$ 

The map $\p^1$ sends the one-form $pdq \in \Omega^1_R$ to the cocycle
$pq \in C^1_f$. The section of the cohomology bundle associated to
$pq\in C^1_f $ is obtained by restricting the class of the form $pdq$ to the fibres of $f$.
The section associated to
$1=pq/pq\in C^1_f $ is obtained by restricting the class of the form $pdq/pq=dq/q$ to the fibres of $f$. The class $[dq/q]$
generates the first de Rham cohomology group of the fibre which is in this case one dimensional. Of course any multiple of
$[dq/q]$, such as $[pdq]$, also generates this group. There is no $H^2(f)$ group for $n=1$ and the problem of quantising an integrable system is in this case empty.
\end{example}
\begin{example}Take $n=2$ and $f_1=p_1q_1,\ f_2=p_2q_2$. The complexes $(\Ct^\cdot_f,\dt)$, $(\Omega_f^\cdot,d)$ have both three terms
$$\xymatrix{\Ot_M \ar[r]^d \ar[d]^-{\p_0=\Id} & \Omega^1_f \ar[d]^-{\p^1} \ar[r]^d& \Omega^2_f \ar[d]^-{\p^2}\\
             \Ot_M \ar[r]^-\dt  & \Ct^1_f \approx \Ot_M^2 \ar[r]^-\dt& \Ct^2_f \approx \Ot_M .}$$  The map $\p^1$ sends
the one-form $p_1dq_1 \in \Omega^1_f$ to the cocycle
$(p_1q_1,0) \in C^1_f$ and  $p_2dq_2$ to $(0,p_2q_2)$. The sections of the cohomology bundle associated to the coboundaries
$(1,0),(0,1) \in C^1_f$ are obtained by restricting the cohomology classes of the
forms $dq_1/q_1$ and $dq_2/q_2$ to the fibres of $f$. The classes  $[dq_1/q_1]$ and $[dq_2/q_2]$
generate the first de Rham cohomology group of the fibre which is in this case of dimension two.\\
The section of the cohomology bundle associated to
$(1,0) \w (0,1) \in C^2_f$ is obtained by restricting the cohomology class of the form $dq_1\w dq_2/q_1q_2$ to the fibres of $f$.
The corresponding class generates the second de Rham cohomology group which is of dimension 1.\\
The integrable system $f=(f_1,f_2)$ is quantisable : just take $F_1=f_1,\ F_2=f_2$. 
We see in this very simple example,
that the obstruction space is non-zero but the integrable system is quantisable. In fact we will show that any anomaly class
and in particular any topological anomaly vanishes on $ C^1_f $. In all non-trivial example known to the authors, the group
$H^2(f)$ is non-zero\footnote{Trivial cases may be obtained by taking a family of functions of two variables $f_l(x,y)$
and to consider the integrable system $(\l_1,\dots,\l_n,f_\l(x,y))$ with Poisson structure $\d_x \w \d_y$.}
\end{example}
Our strategy will consist in proving that topological anomalies vanish.
This will show that the anomaly classes associated to an integrable
system are supported on a proper subset.
In particular, if the cohomology modules $H^2(f)$ are torsion free, then this will prove the
quantisation theorem. By Grothendieck's theorem,
to prove the vanishing of the topological classes it is sufficient to do it in the analytic setting.
The proof is based on a complex version of the standard Darboux-Givental-Weinstein and Arnold-Liouville theorems
\cite{Arnold_Liouville,Liouville,Mineur,Givental_Darboux,Weinstein}.
\subsection{Pyramidal mappings}
To conclude this section let us mention a finiteness result concerning the complex $C_f^\cdot$ in analytic geometry.
These results will not be used in the sequel.
 A {\em lagrangian variety} $L \subset M$ on a symplectic manifold $M$ is a reduced analytic space of pure dimension $n$ such that
the symplectic form vanishes on the smooth part of $L$.
\begin{definition}
A holomorphic map $f:M \to S$ is called an {\em integrable system} if its fibres are $n$-dimensional lagrangian varieties.
\end{definition}
In particular for $M=T^*\CM^n$ and $S=\CM^n$ we recover the definition given in Subsection \ref{SS::def}.
If $f:M \to S$ is an integrable system, the module $H^0(f)$ is generated by the components of $f$:
$$\lb g \in \Ot_{M}(U), \{ f_1,g \}=\dots=\{ f_k ,g \}=0 \rb=f^{-1}\Ot_{S}(f(U)). $$
\begin{definition}A {\em pyramidal map}
$f=(f_1,\dots,f_k):M \to S$ is a stratified involutive map such that for any critical point $x$ lying on a stratum $X_\a$,
the hamiltonian vector fields $v_1,\dots,v_k$ of $f_1,\dots,f_k$ give local coordinates around $x$ on $f^{-1}(f(x)) \cap X_{\a}$, i.e.:
$$T_x(f^{-1}(f(x)) \cap X_\a)={\rm Span}\{ v_1(x),\dots,v_k(x) \}$$
at any point $x$ of a stratum $S$ inside the singular locus of a fibre.
\end{definition}
There is a natural notion of a standard representative of a germ of a pyramidal mapping : it is a pyramidal Stein
representative $f:M \to S$ such that the fibres of $f$ are transverse to the boundary of $M$ and all spheres
centred at the origin are transverse to the special fibre \cite{lagrange}.
\begin{theorem}[\cite{VS,lagrange}] The direct image sheaves of the complex $C^\cdot_f$ associated to a standard representative of a pyramidal integrable holomorphic mapping are coherent and the mapping obtained by restriction to the origin
$$(\RM^k f_*\Ct^\cdot_f)_0 \to H^k(f)=H^k(\Ct^\cdot_{f,0})$$
is an isomorphism.
\end{theorem}
\newpage
\section{Darboux theorems in formal symplectic geometry}
\subsection{The formal Darboux-Givental-Weinstein theorem}
Consider a smooth complex analytic manifold $X$.
Let $Y \subset X$ be a closed analytic subspace  defined by an ideal sheaf $\It$.
Recall that the {\em formal completion of $X$ along $Y$}, denoted $\hat X_Y$ or simply $\hat X$,
is the topological space $Y$ together with the sheaf of ideal
$$\Ot_{\hat X_Y}=\Ot_{\hat X}:=\underleftarrow{\lim}\,\Ot_X/\It^k. $$
We call $Y$ the {\em support} of $\hat X$.
\begin{definition}A {\em formal symplectic manifold $\hat M$} is the completion
of a smooth complex analytic manifold $M$ along a closed smooth lagrangian
submanifold $L \subset M$.
\end{definition}
The symplectic structure on $M$ induces a symplectic structure on the formal symplectic space $\hat M$. The formal symplectic manifold $\hat M$ is called {\em Stein} if its support is a Stein manifold.
\begin{theorem}
\label{T::Weinstein}
Any Stein formal symplectic manifold $ \hat M$ supported on a lagrangian subsmanifold $L$
is symplectomorphic to the formal neighbourhood of the zero section the cotangent bundle $T^*L \to L$.
\end{theorem}
\begin{proof}
We adapt the proof of Weinstein to our context \cite{Weinstein}.\\
As $L$ is lagrangian its normal and cotangent bundle are isomorphic.
As the higher cohomology groups of a coherent sheaf on a Stein manifold vanish,
the analytic space $\hat M$ is isomorphic to the formal neighbourhood $\hat T^*L$ of the zero section in the
cotangent bundle of $L$ \cite{SGA1}.
This means that there exists an isomorphism $\psi$
$$ \xymatrix{\hat M \ar[r]^-\psi & \hat T^*L. }$$
Denote by $\omega_1$ the image of the symplectic form in $\hat M$ under the isomorphism $\psi$
and by $\omega_1$ the symplectic form on  $\hat T^*L$.  We interpolate these two symplectic forms
$$\omega_t=(1-t)\omega_0+t\omega_1.$$
We show that the deformation of the symplectic structure $\omega_t$ is trivial, that is, we search a one-parameter
family of biholomorphic maps $\varphi_t$ such that
$$\varphi_t^* \omega_t=0. $$
By differentiating this equation with respect to $t$ and acting by the inverse of the map $\varphi_t^* $, we get the equation
\begin{equation}
\label{E::1}
L_{v_t} \omega_t +\d_t \omega_t=0
\end{equation}
and $v_t$ should be a locally hamiltonian vector field tangent to the manifold $L$.\\
We assert that the 2-form $\d_t \omega_t$ is exact.\\
There is an isomorphism
$$\HM^\cdot(\hat T^*L, \Omega^\cdot_{\hat T^*L}) \approx \HM^\cdot(L, \Omega^\cdot_{L}). $$
The restriction of $\omega_t$ to $L$ vanishes, therefore the class of $ \d_t \omega_t$
in $ \HM^2(L, \Omega^\cdot_{L})$ also vanishes thus  $ \d_t \omega_t$ is exact. This proves the assertion.\\
Write  $ \d_t \omega_t=d \b_t$, there exists a unique vector field $v_t$ such that $i_{v_t}\omega_t=\a_t$ and this vector field satisfies
Equation (\ref{E::1}). This proves the theorem.
\end{proof}
\subsection{The formal Arnold-Liouville theorem}
Any morphism induces a morphism from the formal completion along any of its fibres.
There is a natural notion of an {\em integrable system} $f:\hat M \to \hat S$ on a formal symplectic manifold : it is the restriction of an integrable system to the
formal neighbourhood of a fibre. We shall consider only restrictions to smooth fibres.\\
We shall use the following relative variant of the Theorem \ref{T::Weinstein}.
\begin{theorem}
\label{T::Darboux}
Let $f:\hat M \to \hat S$ be a smooth integrable system. If $\hat M$ is a formal Stein manifold then the map
$f$ is conjugate by a symplectomorphism to a trivial fibration, i.e., there exists a commutative diagram
$$ \xymatrix{\hat M \ar[r]^-\varphi \ar[d]^f & \Spec(\CM[[t]]) \times L \ar[d]^\pi\\
               \hat S \ar[r]&  \Spec(\CM[[t]])}$$
where $\varphi$ is a symplectomorphism, $\pi$ is the projection, $t=(t_1,\dots,t_n)$ and $L$ is the support of $\hat M$.
\end{theorem}
\begin{proof}
The analytic space $\hat M$ is isomorphic to the product $ \Spec(\CM[[t]]) \times L$ and there exists a commutative
diagram
$$ \xymatrix{\hat M \ar[r]^-\psi \ar[d]^f & \Spec(\CM[[t]]) \times L \ar[d]^\pi\\
                \hat S  \ar[r]&  \Spec(\CM[[t]])  }$$
where $\psi$ is an isomorphism \cite{SGA1}.\\
The isomorphism $\psi$ induces a symplectic form $\omega_1$ on the space\\  $T:= \Spec(\CM[[t]]) \times L$
that we interpolate with the given symplectic structure $\omega_0$ of
 $T$:
$$\omega_t=(1-t)\omega_0+t\omega_1.$$
A one-parameter
family of biholomorphic maps $\varphi_t$ such that
$$\varphi_t^* \omega_t=0. $$
exists provided that there exists a locally hamiltonian field $v_t$ tangent to the fibres of $\pi$ such that
\begin{equation}
\label{E::2}
L_{v_t} \omega_t +\d_t \omega_t=0.
\end{equation}
We assert that the 2-form $\d_t \omega_t$ restricted to $\hat M$ is exact.\\
There is an isomorphism
$$\HM^\cdot(T, \Omega^\cdot_{\pi}) \approx \HM^\cdot(L, \Omega^\cdot_L)\otimes \CM[[t]]. $$
The form $\omega_t$ vanishes on the fibres of $\pi$, therefore the cohomology class of  $ \d_t \omega_t$ in
$\HM^\cdot(L, \Omega^\cdot_L)\otimes \CM[[t]]$ vanishes. This shows
that the 2-form  $ \d_t \omega_t$ is exact and proves the assertion.\\
Write  $ \d_t \omega_t=d \b_t$, there exists a unique hamiltonian vector field $v_t$ such that $i_{v_t}\omega_t=\a_t$ and this vector field satisfies Equation (\ref{E::2}). This proves the theorem.
\end{proof}
\begin{remark} In this theorem the symplectomorphism is in general not analytic but only formal. To see it, consider the pencil of plane cubics
$$C_a=\{(x,y) \in \CM^2: y^2+x^3+ax+a=0 \}. $$ 
The modulus of the corresponding compactified curve varies according to the value of $a$ and therefore the affine curves $C_a$
are not isomorphic as complex analytic manifolds, the family is a fortiori symplectically non-trivial.
\end{remark}
Theorem \ref{T::Darboux} has the following corollary which might be regarded as a complex version of the Arnold-Liouville theorem.
\begin{theorem}
\label{T::AL}
Let $f=(f_1,\dots,f_n):\hat M \to \hat S$ be a smooth Stein formal integrable system, then
there exists commuting vector fields $Y_1,\dots,Y_n $
which commute with the hamiltonian fields of the component of $f$ and such that $\omega(X_i,Y_j)=\dt_{ij}$.
\end{theorem}
\begin{proof}
The previous theorem implies that we may restrict ourselves to the case where $M$ is a product $T^*L=\CM^n \times L$
and $f$ is the projection to $\CM^n$. The choice of this model allows us
to fix the vector fields $Y_1,\dots,Y_n$ by requiring that they should be tangent
to the fibres of the projection $T^*L \to L$. If we take local coordinates $(q_1,\dots,q_n)$ in $L$ so that the symplectic
structure is given by $\sum_{i=1}^n dt_i \w dq_i$
then the vector fields $Y_1,\dots,Y_n$ are the hamiltonian fields $\d_{t_1},\dots,\d_{t_n}$ of the projection to $L$.
\end{proof}
\begin{definition} We call a set of pairwise commuting vector fields $X_1,\dots,X_n,$ $Y_1,\dots,Y_n$ on a complex symplectic manifold
such that  $\omega( X_i,Y_j )=\dt_{ij} $ a set of {\em action-angle vector fields}. 
\end{definition}
\section{Quantisation of formal integrable systems}
\subsection{Action-angle star products}
To a set $\s=\{X_1,\dots,Y_n\}$ of action-angle vectors fields on a formal symplectic manifold,
we associate the {\em action-angle star product}
$$f \star_\s g=\sum_{i,k} \frac{\hbar^k}{k!}X_i^kfY_i^kg $$
where $X^k\cdot$ and $Y^k\cdot$ denote respectively the $k$-th Lie derivative
along $X$ and $Y$.
\begin{theorem}
\label{T::Fedosov}
The action-angle star products on a formal Stein symplectic manifold $(\hat M,\omega)$ are all equivalent, i.e.,
they define isomorphic sheaves of non-commutative algebras.
\end{theorem}
\begin{proof}
In the $C^\infty$ case these star products would be Fedosov products associated to a flat symplectic connection on $(\hat M,\omega)$ with Weyl curvature form $-(i/\hbar)\omega$ and are therefore equivalent \cite{Fedosov}, Theorem 4.3 (see also \cite{Weinstein_star} and \cite{Kontsevich_Poisson} for the more general case of a Poisson manifold).
The holomorphic case is similar to the $C^\infty$ case. We give an outline of the proof in our situation.\\
A set of action-angle vector fields $\s=\{X_1,\dots,Y_n\}$ defines a flat connection on $\hat M$
for which the vector fields are horizontal sections.\\
The  normal product on each linear space $T^*_x M  $ induces a product in $\Ot_{T^* M}[[\hbar]]$ that we denote by $\star_F$. For instance if $M=\CM^2=\{(x,y)\}$ and $T^*M=\CM^4=\{(x,y,\xi,\eta)\}$ then
the only non-commutative product among linear forms is $\eta \star_F \xi=\xi\eta+\hbar $. In this case the only non-trivial commutator
is $[\eta,\xi]=\hbar$.\\
The flat connection on $M$
induces an isomorphism of a symmetric neighbourhood $V \subset M \times M $ of the diagonal with a neighbourhood $V'$
of the zero section in $TM$. Here symmetric means invariant under the involution
$(a,b) \mapsto (b,a) $. The symplectic form in $M$ induces an isomorphism of $TM$ with $T^*M$. In the sequel, we make this identifications.\\
Using the flat connection, we associate to a vector field $v:M \to V' \subset TM $ is associated two vector fields
$v',v'':TM \to T(TM) \approx TM\times TM$ with coordinates $(v,0)$ and $(0,v)$ in the decomposition of $T(TM)$ into horizontal and vertical components.\\
Given an open subset $U \subset M$, we define the subalgebra
$\G_\s(U) \subset \G((U \times U) \cap V',\Ot_{T^*M}[[\hbar]],\star_F)$ by
\begin{equation}
\label{E::Fedosov}
f \in \G_\s(U) \iff X_i'f=X_i''f,\  Y_i'f=Y_i''f \ \forall i .
\end{equation}
\begin{lemma} The algebra $(\G_\s(U),\star_F)$ is isomorphic to the algebra $(\G(U,\Ot_M[[\hbar]]),\star_\s)$, i.e.,
for any $f_0 \in \G(U,\Ot_M[[\hbar]])$ there exists a unique $f \in \G_\s(U)$ satisfying the system of equation (\ref{E::Fedosov})
and the map $f_0 \mapsto f$ induces an isomorphism of algebras.
\end{lemma}
\begin{proof}
To get a clear geometrical picture, let us take the times $x_1,\dots,x_n$,\\ $
y_1,\dots,y_n$ of the vector fields $X_1,\dots,X_n$, $Y_1,\dots,Y_n$ as local coordinates.\\
The tangent bundle has coordinates $x_1,\dots,y_n $ and $\xi_1,\dots,\xi_n,\eta_1,\dots,\eta_n$
and in these local coordinates we have
$$X_i'=\d_{x_i},\ Y_i'=\d_{y_i},\ X_i''=\d_{\xi_i},\ Y_i''=\d_{\eta_i}. $$
So, the subalgebra $\G_\s(U)$ consist of functions $f$ such that
$$\d_{x_i}f=\d_{\xi_i}f,\ \d_{y_i}f=\d_{\eta_i}f. $$  
For each $f_0 \in \G(U,\Ot_M[[\hbar]])$ there is a unique $f$ satisfying this system of partial differential equations for which
$f_{\xi=0,\eta=0}=f_0$, namely
$$f(x,y,\xi,\eta)=f_0(x+\xi,y+\eta). $$
The isomorphism of both algebras is immediate. For instance
we have
$$y_j \star_\s x_i=y_j x_i+\hbar \dt_{ij} .$$
The linear forms $y_j,x_i \in \G(U,\Ot_M[[\hbar]]) $ correspond to the linear forms
$ y_j+\eta_j,x_i+\xi_i \in \G_\s(U)$.
The corresponding normal product is
$$ (y_j+\eta_j) \star_F (x_i+\xi_i)= (y_j+\eta_j)(x_i+\xi_i)+\hbar \dt_{ij} $$
as $\eta_j \star_F \xi_i=\hbar \dt_{ij}$.
\end{proof}
Given two sets $\s=\{X_{1,\s},\dots,Y_{n,\s}\},\ \s'=\{X_{1,\s'},\dots,Y_{n,\s'}\}$ of action angle vector fields,
there is in each tangent space $T_xM$ a unique map $A_x$ which sends the $i$-th vector in the list $\s$ to that of the list $\s'$. Thus, we get an isomorphism
$$A:TM \to TM,\ A_*X_{i,\s}=X_{i,\s'},\  A_*Y_{i,\s}=Y_{i,\s'}. $$
The map $A$ induces an isomorphism between the algebras $\G_\s(U)$,  $\G_{\s'}(U)$ and therefore of the 
sheaves $\Ot_M[[\hbar]] $ with the star products associated to $\s$ and $\s'$. This concludes the proof of the theorem.
\end{proof}

\subsection{The lifting property}
\begin{proposition}
\label{P::quantisation1}
Let $f=(f_1,\dots,f_n):\hat M \to \hat S$ be a smooth integrable system defined on a Stein formal symplectic manifold $\hat M$.
There exists $\star$-commuting elements $F_1,\dots,F_n \in \G(\hat M,\Ot_{\hat M}[[\hbar]])$ which lift $f_1,\dots,f_n$.  
\end{proposition}
\begin{proof}
Denote by $X_1,\dots,X_n$ the hamiltonian vector fields associated to $f_1,\dots,f_n$.
By Theorem \ref{T::AL} there exists a set of action-angle vector fields $\s=\{ X_1,\dots,X_n,Y_1,\dots,Y_n \}$.
The functions $f_1,\dots,f_n$  commute for the  star product $\star_\s$. By Theorem \ref{T::Fedosov},
there exists an isomorphism of sheaves
$$\p:(\Ot_M(\hat M)[[\hbar]],\star_\s) \to (\Ot_M(\hat M)[[\hbar]],\star)$$
where $\star$ denotes the normal product in $M \subset T^*\CM^n$. The elements $F_1=\p(f_1),$ $\dots,F_n=\p(f_n)$ commute
for the normal product and lift $f_1,\dots,f_n$. This proves the proposition.
\end{proof}
We may be even more precise.
\begin{proposition}
\label{P::quantisation2}
Let $f=(f_1,\dots,f_n):M \to S$ be a smooth integrable system defined on a Stein formal symplectic manifold $\hat M$
and let $G=(G_1,\dots,G_n) \in
\G(\hat M,\Ot_{\hat M}[[\hbar]]/(\hbar^{l+1}))$ be an $l$-lifting of $f$.
 There exists $\star$-commuting elements $F_1,\dots,F_n \in \G(\hat M,\Ot_{\hat M}[[\hbar]])$ which lift $f$ and project to $G$.  
\end{proposition}
Proposition \ref{P::quantisation2} will be reduced to Proposition \ref{P::quantisation1} by showing that all liftings are isomorphic.
Let us now explain in which sense these liftings are isomorphic.\\
 If we compose an integrable system $f=(f_1,\dots,f_n):M \to S$ with a biholomorphic map
$\psi:S \to S'$ we get another integrable system $\psi \circ f$. This procedure can be quantised: consider
an automorphism $\psi=\sum_k a_k z^k \in \Aut(\G(\hat S,\Ot_{\CM^n}[[\hbar]]^n))$ then $\psi \circ F:=\sum a_k F^k$ is also an $l$-lifting of
$f$.
\begin{proposition}
\label{P::quantisation3}
Consider the fibre $L$ of  a smooth Stein lagrangian fibration $f=(f_1,\dots,f_n):M \to S$.
Let $F_1,\dots,F_n,\ G_1,\dots, G_n,\ F_i,G_i \in \G(\hat M,\Ot_{\hat M}[[\hbar]]/(\hbar^{l+1}))$
be two $l$-liftings of the integrable system $f$.
These two liftings are equivalent in the formal neighbourhood $\hat M$ of $L=f^{-1}(s)$, i.e., there exists automorphisms
$\p \in \Aut  \G(\hat M,\Ot_{\hat M}[[\hbar]]/(\hbar^{l+1}))$, $\psi \in \G(\hat S,\CM[[\hbar]])$ such that
$$(\p(G_1),\dots,\p(G_n))=\psi \circ (F_1,\dots,F_n).$$
\end{proposition}
\begin{proof}
Denote by $X_i$ is the hamiltonian field associated to $f_i$.
By Theorem \ref{T::Darboux} and Theorem \ref{T::Fedosov}, we may replace $\hat M$ by $L \times \Spec(\CM[[t]]) $, $f$ by the projection on $L$ and the normal product by the star product associated to
action-angle vector fields $X_1,\dots,Y_n$ with $Y_i=\d_{t_i}$.\\ 
Let us prove that any $l$-lifting $F=(F_1,\dots,F_n)$ of the type
$$F_1=f_1,\dots,F_{j-1}=f_{j-1},F_j=f_j+\hbar^l g_j,\dots,F_n=f_n+\hbar^l g_n$$
is equivalent to an $l$-lifting $F'=(F_1',\dots,F_n')$ of the type
$$F_1'=f_1,\dots,F_{j-1}'=f_{j-1},F_j'=f_j, F_{j+1'}=f_{j+1}+\hbar^l g_{j+1}',\dots,F_n'=f_n+\hbar^l g_n'$$
Then by induction on $j$ and $l$ this will prove that any lifting is equivalent to the trivial lifting $f_1,\dots,f_n$ and will conclude
the proof of the proposition.\\
The symplectic form induces an isomorphism between one-forms and vector fields.
Let $\p^t$ be the flow of the vector field $v$ associated to the one-form $\a=-g_j dt_j$ at time $t$. The quantisation of $\p^t$
evaluated at $t=\hbar^l$ gives an automorphism which maps $F$ to $F'$. This can be seen locally : the restriction of the
one form $\a$ to a contractible open subset is exact $\a=dH$ and $g_j=-\{H, f_j \}=-L_{X_j} H$, that $F$ is an $l$-lifting
implies that $\{ H,f_i \}=0$ for $i<j$.
The automorphism $\p^t$, $t=\hbar^l$, is defined by
$$A \mapsto \exp(\hbar^{l-1}H)\, A \,\exp(-\hbar^{l-1}H).  $$
This automorphism depends only on $dH$ and not on the choice of $H$.
It maps $F_i'$ to $F_i'+\hbar^{l}\{ H,f_i \}=f_i$ for $i\leq j$. 
This concludes the proof of the proposition.
\end{proof}
\subsection{Proof of theorem \ref{T::quantisation}}
By Proposition \ref{P::quantisation2} all topological anomalies vanish.
Therefore the anomaly classes are supported on the singular values of $f$.
If  $H^2(f)$ is a torsion free module, there is no section of the sheaf supported on a proper subset.
This shows that any anomaly class vanishes and concludes the proof of the theorem.
\bibliographystyle{amsplain}
\bibliography{master}
\end{document}